\documentclass[12pt,preprint]{aastex}
\slugcomment{draft version}
\usepackage{graphicx,times}
\usepackage{longtable}

\shorttitle{The properties of cross-correlation and spectra  of   4U 1608-52}
\shortauthors{Y.J. Lei et al.}

\begin{document}

\title{ The properties of cross-correlation and spectra  of the low-mass X-ray binary 4U 1608-52}

\author{Ya-Juan Lei\altaffilmark{1},  Shu Zhang\altaffilmark{2}, Jin-Lu Qu\altaffilmark{2}, Hao-Tong Zhang\altaffilmark{1}, Zhi-Bing Li\altaffilmark{3}, Cheng-Min Zhang\altaffilmark{1},
Yong-Heng Zhao\altaffilmark{1}}

\altaffiltext{1}{Key Laboratory of Optical Astronomy, National Astronomical Observatories, Chinese
Academy of Sciences, Beijing 100012, China; leiyjcwmy@163.com}

\altaffiltext{2}{Particle Astrophysics Center, Institute of High
Energy Physics, Chinese Academy of Sciences, Beijing 100049, China}

\altaffiltext{3}{Xinjiang Astronomical Observatory, Chinese Academy of Sciences, 150, Science 1-Street, Urumqi, Xinjiang 830011, China}

\begin{abstract}

With {\it RXTE} data, we analyzed the cross-correlation function between the soft and  hard
X-rays of the transient atoll source 4U 1608-52. We found anti-correlations in three outbursts occurred in
1998, 2002 and 2010, and significant time lags of several hundreds of seconds  in the latter two outbursts.
 Our results show no correlation between the soft and hard X-rays in the island state, and a dominated positive correlation
 in the lower banana state. Anti-correlations are presented at the upper banana state for the outburst of 2010 and
at the lower left banana states for the other two outbursts. So far for atoll sources the cross-correlation has been
studied statistically only for  4U 1735-44, where anti-correlations showed up in the  upper banana state. Here our investigation upon
4U 1608-52 provides a similar result in its 2010 outburst.
In addition, we notice that the luminosities in the upper banana of of 1998 and 2002 outbursts are about 1.5 times
that of 2010 outburst whose luminosity in the upper banana is close to that of 4U 1735-44.
The results suggest that the states in color-color diagram of a source could be correlated with the luminosity of the source.
A further spectral analysis shows that, during the 2010 outburst,
although an anti-correlation presents at the highest fluxes, the contemporary spectrum is not the softest one along the outburst evolution.
This suggests that the observed anti-correlation may be relevant to the transition between the hard and soft states,
which is consistent with the previous results on 4U 1735-44 and several black hole X-ray binaries that anti-correlations are observed during the transition states.

\end{abstract}


\section{Introduction}

Low-mass X-ray binaries (LMXBs) with a neutron star (NS) primary can 
be divided into two subclasses, atoll and Z sources, according to their X-ray spectral and timing properties which 
depend on the position of the source in the color-color
diagram (CCD) \citep{has89}.
It is suggested that the accretion rate and  magnetic field are different for  atoll and
Z sources \citep{zha07}.
The luminosity   is about  0.001 - 0.5 $L_{Edd}$ for atoll  sources, while   0.5-1 $L_{Edd}$ for Z sources
\citep[e.g.,][]{for00, van06}. Also the magnetic field is low for atoll sources ( $B \sim 10^{8} G$), and
  high for Z sources ( $B > 10^{9} G$) \citep{has89}. An obvious morphology difference in  CCD is that,
  the typical atoll source forms a C-shaped track  \citep[e.g.,][]{dis01}, whereas the Z source traces out a Z track \citep[e.g.,][]{jon00}.

The atoll sources contain the extreme island state (EIS), the island state (IS),
and the banana branch, which is usually subdivided into lower left banana
(LLB), lower banana (LB), and upper banana (UB) states.
The movement of an atoll source from IS to UB is accompanied with an increasing luminosity which usually denotes an increasing accretion rate $\dot{M}$.
The Z source track in the
CCD  is divided into three main branches: horizontal branch (HB), the normal branch (NB) and flaring branch (FB).
The accretion rate is assumed to increase along the direction of HB-NB-FB \citep{hasetal90}.
However, recently, the results of studying the source XTE J1701-462
suggest that the branches of the Z track  might correspond
to a roughly constant accretion rate \citep{lin09, hom10}. In
addition, \citet{jac09} analyzed the spectra of the Z source GX 5-1, and
indicated that the mass accretion rate at the soft apex  reaches its minimum.
For  Z sources, the
low frequency quasi-periodic oscillations (QPOs) of $\sim$5-$\sim$60 Hz and Kilohertz (kHz) QPOs are usually detected on the HB and
upper NB \citep[e.g.,][]{van00, cas06, van06}.

Usually, there are three models that have been used to explore the spectral  properties
of NS LMXBs: eastern, western and  hybrid models.
In the Eastern model, a  multicolor disk blackbody (MCD) and a weakly Comptonized blackbody 
are used for describing the thermal and Comptonized components \citep[e.g.,][]{mit89, agr03}. 
 In the Western model, a single-temperature blackbody (BB) from the boundary
layer represents the thermal
component and Comptonized emission is from the accretion disk \citep[e.g.,][]{whi88, chu95}. 
\citet{lin07} suggested a hybrid model,
BB plus a broken power law (BPL) are used for the hard state, and
two thermal components (MCD and BB) plus a
constrained BPL are suit for the soft state.
In present, the hybrid model is the most proper for NS LMXBs.
At low luminosities, the spectra of atoll sources are hard, which are similar to the low/hard of black hole X-ray binaries (BHXBs) \citep[e.g.,][]{van94}.
and at high luminosities, the spectra are much soft \citep[e.g.,][]{dis00}.
However, the spectra of Z source are always soft during its three branches.

Both the spectral and timing  analyses  are approaches for understanding physical changes in the sources emission.
The evolution of cross-correlation function between different X-ray energy bands is a useful means  for
exploring  geometrical structure of the accretion disk.
Earlier works show already detection of  anti-correlations for Z sources and BHXBs. 
The cross-correlations of two Z sources (Cyg X-2, GX 5-1) have been analyzed and  anti-correlated lags of $\sim$ several 10-100 s are found  in the HB and upper
NB \citep{lei08, sri12, lei13a}.
BHXBs  consist of three states, hard, thermal, and steep
power law (SPL) states, where SPL state represents the intermediate state \citep{mcc06}.
Such anti-correlated lags of a few hundred seconds are
also detected in the SPL states of a few  BHXBs \citep{cho05, sri07, sri09, sri10}.
The large timescale of lags could be
interpreted  by  the viscous timescales of the inner accretion disk \citep{sri12}.
\citet{sri10} suggested in their simulations that the observed phenomena of an anti-correlation should be relevant to
 some real physical processes,  of which we know little from the current observations.

A tip of the iceberg of having anti-correlations for atoll sources was firstly revealed in 4U 1735-44 (Lei et al. 2013a), and then on
4U 1608-52 from one outburst \citep{lei13b}. To probe the whole, it is obvious that one needs more observations to consolidate the
results.
In the paper, we  study the cross-correlations of 4U 1608-52 statistically with all   available  {\it RXTE} data .
 4U 1608-52 is a transient source that undergoes outburst with a waiting  period  of varying from 70 days to years,
 and a flux variability of  about 2 orders of magnitude.
4U 1608-52 has a distance of 5.8 kpc, neutron star mass of
M = 1.74 $\pm$ 0.14$M_{\odot}$  \citep{guv10}.
Its X-ray spectra have been studied in some works \citep[e.g.,][]{gie02, che06, lin07}.
The timing properties of 4U 1608-52 have been analyzed, and kHz QPOs have
been detected  with frequencies ranging from $\sim$ 500
to $\sim$ 1050 Hz \citep[e.g.,][]{ber96, men98, van03}.
\citet{vau97, vau98} and \citet{ave13} show the energy and frequency dependence of the Fourier time lags and intrinsic
coherence of the kHz QPOs.
 The organization of the paper is that,  the observations and data analysis of {\it RXTE} are introduced in Sec.2,
 the results derived from investigating the evolution of the cross-correlation between the
soft (2-3.3 keV)  and hard (12-30 keV) X-rays along the CCD are shown in Sec.3,
and these findings are compared with those from 4U 1735-44, Z sources and a few BHXBs in Sec.4.

\section{Observation and Data Reduction}

The long-term light curves of 4U 1608-52 from the RXTE All-Sky Monitor (ASM)
are shown in Figure 1 (Levine et al. 1996). Embedded in light curves are a variety of outbursts showing up with different morphology in
   amplitude and duration.
We analyzed  all data available from the {\it RXTE} proportional
counter array (PCA)   observations, covering a time period of    1996
March 3 -- 2011 December 28. PCA contains 5 non-imaging, co-aligned Xe multiwire proportional
counter units (PCUs) \citep{jah06}.
We exclude data of those with earth limb elevation angle  less than $10^{o}$ , pointing offset greater
than $0.02^{o}$, or  influence by the type I bursts.
Only PCU2 data are adopted because PCU2 is the best calibrated unit and has the
longest observational duration.
The light curves  are extracted from the data of standard 2 mode  with bin size of 16 s, and prepared for carrying out
the cross-correlation function after excluding those with duration shorter than 2000 s.
The cross-correlation function is obtained with the XRONOS tool ``{\it crosscor}'' which computes the cross correlation by a FFT algorithm.
With this tool, the coefficient and time lag are estimated simultaneously  between different energy bands
 \citep{bri74, wei75}. In this work we take  two energy bands, defined as the soft at 2-3.3 keV and the hard at 12-30 keV.
%

The cross-correlation results are divided into three groups in the same way of the previous work:
positive, ambiguous and anti-correlated cases \citep[also see][]{lei08}.
The correlated coefficient and  time lag  can be obtained by  fitting  the  anti-correlated part of the  cross-correlation function with an
inverted Gaussian function.
For the observations detected with anti-correlation,
Figure 2 shows the background subtracted light curves, the anti-correlations between soft and
hard X-rays, and the pivoting spectra.
 The pivoting spectra  are obtained from the hard regions (hardness ratio being 10\%  more than the average at  12-30
keV/2-3.3 keV) and the soft regions (hardness ratio being 10\%  less than the average at  12-30
keV/2-3.3 keV).

For building the CCDs and studying the distribution of cross-correlated results,  
the soft and hard colors are defined  as the count-rate ratios 3.5-6.0
keV/2.0-3.5 keV and 9.7-16 keV/6.0-9.7 keV \citep[also
see][]{obr04}. The background subtracted light curves are extracted from standard 2 mode data at  2.0-3.5 keV,
3.5-6.0 keV, 6.0-9.5 keV and 9.5-16 keV, respectively.
Figure 3 shows the CCDs of the outbursts where the anti-correlations are found in  512 s time bins. For spectral
analysis, the PCA background is subtracted  with the
latest versions of the appropriate background models for Epoch 3 of
the mission (1996 April 15 -1999 March 22).  A  systematic
error of 0.6\% is added to the spectrum to account for the calibration
uncertainties in spectral fitting with  software $XSPEC$ version
12.8.0 \citep{sch91, arn96}.

\section{Results}

\subsection{The results of cross-correlations  and  their  distribution on the CCD}

With all the data  of {\it RXTE},  for each segment of the light curves of the transient atoll source 4U 1608-52, 
we have analyzed the cross-correlation  function between the soft (2-3.3 keV) and  hard (12-30 keV) X-rays.
The anti-correlations are detected in these outbursts occurred in 1998, 2002 and 2010, and the results are listed
 in Table 1 for those observations with  positive or ambiguous correlations,
 their average hardness ratios and the time lag estimated from the positive correlations.
One sees from  Figure 3 that, for the 1998 and 2002 outbursts,   ambiguous correlations are for all IS and dominant for
 LLB,  but the positive correlations show up in LB by $>$ 50 percents.
For the 2010 outburst, we have both positive and anti-correlations in the UB, and the results in  LLB and LB are similar to the other two outbursts.
We notice that, the time lags are detected in general insignificant for all the three outbursts. The most significant detection is
at about 3 $\sigma$ level, derived from one observation in LLB of 1998 outburst (see Table 1 and Figure 2),
with a positively correlated lags of several hundred seconds.

The anti-correlations are detected from two observations in 1998 outburst,
one observation in 2002 outburst and one observation in 2010 outburst, as shown in Figure 2 for their
 light curves and anti-correlations. The corresponding results are presented in Table 2   and 
Figure 3. We notice the different locations in CCD for the detected anti-correlations: in LLB
(i.e., the transient state between IS and LB) for the 1998 and 2002 outbursts and in UB for the 2010 outburst.
The time lags are not obvious in the 1998 outburst but significantly presenting in 2002 and 2010 outbursts.
The hard X-rays are found to lag to soft X-rays by 168 $\pm$ 32 s on October 1, 2002, and 628 $\pm$ 43 s on March 8, 2010.
Similar results were reported previously from the  the atoll source 4U 1735-44, the Z sources Cyg X-2
and GX 5-1 \citep{lei08, sri12, lei13a}. We show in Figure 2 the spectral evolution of
 the soft and hard regions of the light curves and in Table 2 the pivoting energies.

\subsection{ Spectral  variation }

Both the spectrum and the cross-correlation function are strongly correlated with the position in the CCD.
For understanding the spectral dependence of the cross-correlation function, we divide the CCD of the 2010 outburst into 8 regions,
as shown in the right panel of  Figure 3, and perform spectral fitting for the data clumped within each region.
 Based on the so-called  hybrid model of \citet{lin07}, we fit these spectra at 2.5-25 keV with the multi-color disk model ($diskbb$
in XSPEC), blackbody ($bbody$ in XSPEC) plus broken power-law ($bknpower$ in XSPEC).
The iron line is fitted by a Gaussian function with a center energy  constrained within 6.2-7.3 keV and a width fixed at 0.1 keV \citep{lin07}.
The photo-electric absorption
column ($wabs$ in XSPEC) is fixed at the value of $N_{\rm H}$ = 1.0 $\times$ $10^{22}$
cm$^{-2}$ \citep{pen89}, since the low energy coverage makes it  hard to estimate the absorption with PCA observation alone.
In practice of the spectral fitting with BPL, the broken energies are constrained to $>$ 10 keV and the initial photon index  to
$<$ 2.5 \citep{lin07}. Figure 4 shows the fitted spectra.

The spectral fitting results are listed in Table 3. The source is in LLB for regions 1-3,
where the Comptonized components are found to contribute to about $\sim$ 20\% - 50\% \citep{lin07}.
Along with the flux increasing in regions 4-7, the soft X-ray components (disk blackbody and blackbody)  start to play a role,
which is in accordance with an increasing  accretion rate in CCD.
Both the temperatures of the disk and blackbody increase with the flux till the region 7,
beyond of which the temperature starts to decrease in region 8 where the flux reaches to a maximum.
This is similar to that observed in  4U 1735-44: the spectrum of the highest flux is not the softest, which suggests undergoing a state transition.

%
%

%
\section{Discussions}

\subsection{ Comparison to the atoll source 4U 1735-44}

For probing the cross-correlated evolution of atoll sources, which was hinted firstly from 4U 1735-44,
we analyzed the cross-correlations of X-ray transient atoll source 4U 1608-52 systematically,
and found results similar to those from 4U 1735-44: the ambiguous correlation dominates the IS and LB
are dominated respectively by correlations of the ambiguous and the positive.
In general, the cold photons are supposed to come from the accretion disk, and the high energy photons  from the hot corona.
In the IS, the low accretion rate leads to the accretion disk boundary be blocked  away from the hot corona,
and hence the ambiguity  soft and hard X-rays. In the LB, the  accretion rate increases and the accretion disk moves inward.
The closer corona and inner disk favor a response of the corona to the disk seed photons and hence a positive correlation between the soft and hard X-rays.

There are 4 observations with anti-correlation detected through 1998, 2002 and 2010 outbursts, among them 2 show  significant time lags.
For the 1998 and 2002 outbursts, the anti-correlations are detected in the LLB, which are corresponding to
the transitions between the hard state (IS) and soft state (LB).
For the 2010 outburst, the anti-correlation is detected in the top of UB, which
corresponds to the maximum flux for the outburst.
According to the results of  \citep{lin07}, for atoll sources, at the highest flux,
the Comptonized component is at the lowest and   negligible level.
From the results of spectral fitting of 2010 outburst, in the top of UB where anti-correlation is detected, i.e., at the highest flux,
the Comptonized component is not the lowest. Therefore, the top of UB could be corresponding to the transition states.
These results of 4U 1608-52 are consistent with the results of atoll source 4U 1735-44 whose anti-correlation are also detected in the UB.
In addition, we notice that the highest luminosity of 2010 outburst is close to that of 4U 1735-44,
and the highest luminosities of 1998 and 2002 outbursts are nearly 1.5 times higher than that of 2010 outburst.
Therefore, the position of anti-correlation could be related to the range of the luminosity, which
could be affected by the geometry and physical condition of the accretion disk.
When the highest luminosity of the outburst is large, the anti-correlation occurs in LLB, and vice versi, the anti-correlation occurs in UB.
The anti-correlation and luminosity could have a certain correlation.
These results suggest that the transition states in CCD of a source may be  in  evolution  with the luminosity of the source.
For exploring the relation, more sources need to be studied.

\subsection{ Comparison to Z sources (Cyg X-2 and GX 5-1)}

There are two Z sources (Cyg X-2, GX 5-1) whose cross-correlations have been studied,
and the anti-correlations were detected in the HB and the upper NB \citep{lei08, sri12}.
However, the anti-correlated observations of atoll sources are found in LLB or UB.
For atoll source, the accretion rate increases from IS to UB. Therefore, the relation between the accretion rate and
the cross-correlation of soft and hard X-rays is relatively clear.
For Z sources, the relation between the accretion rate and the position of CCD is controversial.
Assuming  that the accretion rate increases along the HB-NB-FB on the CCD, then the anti-correlations should be corresponding to the lower accretion rate.
However, \citet{jac09}'s  model for Z track suggested that the accretion rate is low in the soft apex between NB and FB,
in this case, the anti-correlations are consistent with the high accretion rate.
There's also the perception that the accretion rate could be constant along the Z track \citep{lin09}.
For Z sources, the distribution of cross-correlation in the CCD is known, but the relation between it and the accretion rate is still an open question.

For Z sources, a movement along  CCD  is in general  accompanied with different evolution of radio activities.
The radio emission become stronger in the HB. Therefore, the structure of accretion disk of Z sources could be
correlated with the jet where radio emissions come from. Most of atoll sources usually show  radio luminosity lower than  Z sources,
thus our understanding  is still rather vague.
\citet{mig03}'s studies about atoll source 4U 1728-34 show that, weaker and persistent radio emission is observed
when the source is steadily in the hard X-ray state,
 and  the strongest and most variable emission seems to be associated with
the transitions between hard (IS) and softer (LB) X-ray states.
If a similar situation is also held for 4U 1608-52, the anti-correlations that are detected in the transient states may be relevant to
with radio flaring, which in turn suggests that the structural changes of the accretion disk could be  coupled with the jet activity.

\subsection{ Comparison to some BHXBs in a context of the truncated disk model}

Based on the timing and spectral characteristics, BHXBs can be divided into five distinct states:
quiescent state, low/hard state(LHS), intermediate state (IMS), high/soft state
(HSS), and very high state (VHS) \citep{rem05, rem06, bel10}.
Due to similar spectral and timing  features, the VHS is also taken as IMS,
and represents the transitions between LHS and HSS \citep{don07}.
Atoll sources share lots of properties with BHXBs.
A few BHXBs were reported with several 100-1000 s anti-correlated lags in the CCF
of soft and hard X-ray bands in the SPL/IM states \citep{cho05, sri07, sri09, sri10}.
We find for atoll source 4U 1608-52 anti-correlations  in LLB and UB. Similar to those in  the atoll source 4U 1735-44,
for 4U 1608-52, the LB could be responsible to the HSS of a black hole, and the UB  to the VHS.
Therefore, our results of atoll sources are similar with those of BHXBs: both of their anti-correlations are detected in the transition states.

For BHXBs, the truncated accretion disk model is popularly used to explain the
spectral and temporal features \citep{don07}.
It is proposed that an accretion flow of BHXB consists of two zones with a transition radius,  an inner
advection-dominated accretion flow (ADAF) and an outer standard thin
disk \citep{esi97}.
The disk is usually considered to be truncated
at a large radial distance in the LHS,  and  it becomes
non-truncated in the  HSS.
For the SPL/IM state, the spectral
and temporal results of the studying various black hole sources
suggest that the accretion disk is truncated in a radius very
close to the black hole \citep[e.g,][]{sri10}.
In the BHXBs, the detected anti-correlation indicates that
the  ADAF condenses and expands to the inner disk,  resulting in that the anti-correlation in flux between the soft and hard X-rays.
 The IMS has been considered as the most probably  spectral state  in which the condensation
of the inner hot matter can transform into an inner disk
\citep{liu07, mey07, sri10}.
However, the truncated accretion disk is not supported by some observational results \citep{mil06, ryk07}.
Analysis of iron line is helpful for understanding the structure and physics of the innermost accretion disk \citep{ste90}.
 \citet{don06} suggested
that the detected smeared  iron emission line can be explained  with truncated disk models.
In addition,   the instead narrow iron line of GX 339-4 is consistent with the truncated disk geometry \citep{don10}.
In  the high/soft states of the atoll source 4U 1705-44, a similar profile of iron emission line was found,
which suggests that the accretion disk could be also truncated in the high/soft states of atoll source \citep{dis09}.
The alternative view is that the narrow iron line might also come from the outflows in BHXBs \citep{lau07}.
The physical condition of NS is more complex, our study about anti-correlation is helpful for exploring the geometry of the accretion disk in LMXBs.
The spectral and timing characteristics derived for the different states suggest that the accretion disk is dynamically changing.
The anti-correlated lags of  several hundred seconds can not be explained by only the Comptonization process which only
accounts for $\lesssim$ 1 second timescale lags \citep{has87, now99, bot99}.
The timescale of the anti-correlated lag corresponds to the viscous timescale of the inner accretion disk,
which suggests the viscous re-adjustment timescale of a hypothetical truncation radius.
The observed time lag and spectral changes sustain that the  truncated accretion disk is the most probable physical configuration \citep{sri12}.

The NS systems are different from the BHXBs, which is complex due to  a solid surface  and magnetic field of compact object.
In LMXBs, the hard X-ray emission is usually considered to origin from the Comptonization
of the soft seed photons.
The soft seed photons could origin from the accretion disk or jets in BHXBs \citep{mar04}.
However in NS systems, both the surface of the NS and the accretion disk can provide the soft seed photons.
The  Comptonized region might correspond to a hot corona, a hot flared inner
disk, or even the boundary layer between the NS and accretion disk
\citep[also see][]{pop01, qu01}.
\citet{wen11} suggested that the magnetosphere of NS LMXB might go inside the innermost stable circular orbit (ISCO) when the luminosity 
 go beyond a critical luminosity, and in this case, the NS systems will follow the same evolutionary pattern in the disk of BHXBs.
If the luminosity is beyond  a critical luminosity, the Comptonizing cloud 
shrinks, as the soft seed photons increases, it cools the Compton cloud and hard flux decreases.
The luminosity of the UB is higher, and the accretion disk can extend down to ISCO, 
which can result in  the anti-correlations of the soft and hard X-rays.
According to our spectral analysis, it can be seen that the inner radius of the accretion
disk decreases at higher accretion rate. 
These results are consistent with BHXBs whose anti-correlations are detected in the SPL/IM state where the accretion
disk could be truncated very close to the compact object
\citep{sri10}.

In summary, our studies of 4U 1608-52 provide further evidences for having  anti-correlations in atoll source apart from those hinted firstly in  4U 1735-44.
The anti-correlations are found to occur in CCD at the regions of the LLB or UB,
which could be decided  by the luminosity.
These results are in align with  those discovered in  BHXBs in a sense that,
these  anti-correlations are responding to the state transitions of the source spectral evolution.
A time lag of several hundred seconds as discovered in the data of anti-correlations is consistent with the re-adjustment timescale of truncation disk.
In short, what we found currently upon 4U 1608-52 may reveal so far only a tip of the  iceberg for the atoll sources in general,
and definitely more observations are expected in future to tell us more.

\acknowledgements This research has made use of data obtained through the
high-energy Astrophysics Science Archive Research Center Online
Service, provided by the NASA/Goddard Space Flight Center.  This
work is subsidized by the Natural Science Foundation of China for
grants NSFC numbers 11303047, 11173034, 11173024, 11133002, 11073021, 11203064.

\clearpage

\begin{figure*}
\begin{center}
\includegraphics[width=12cm,angle=0,clip]{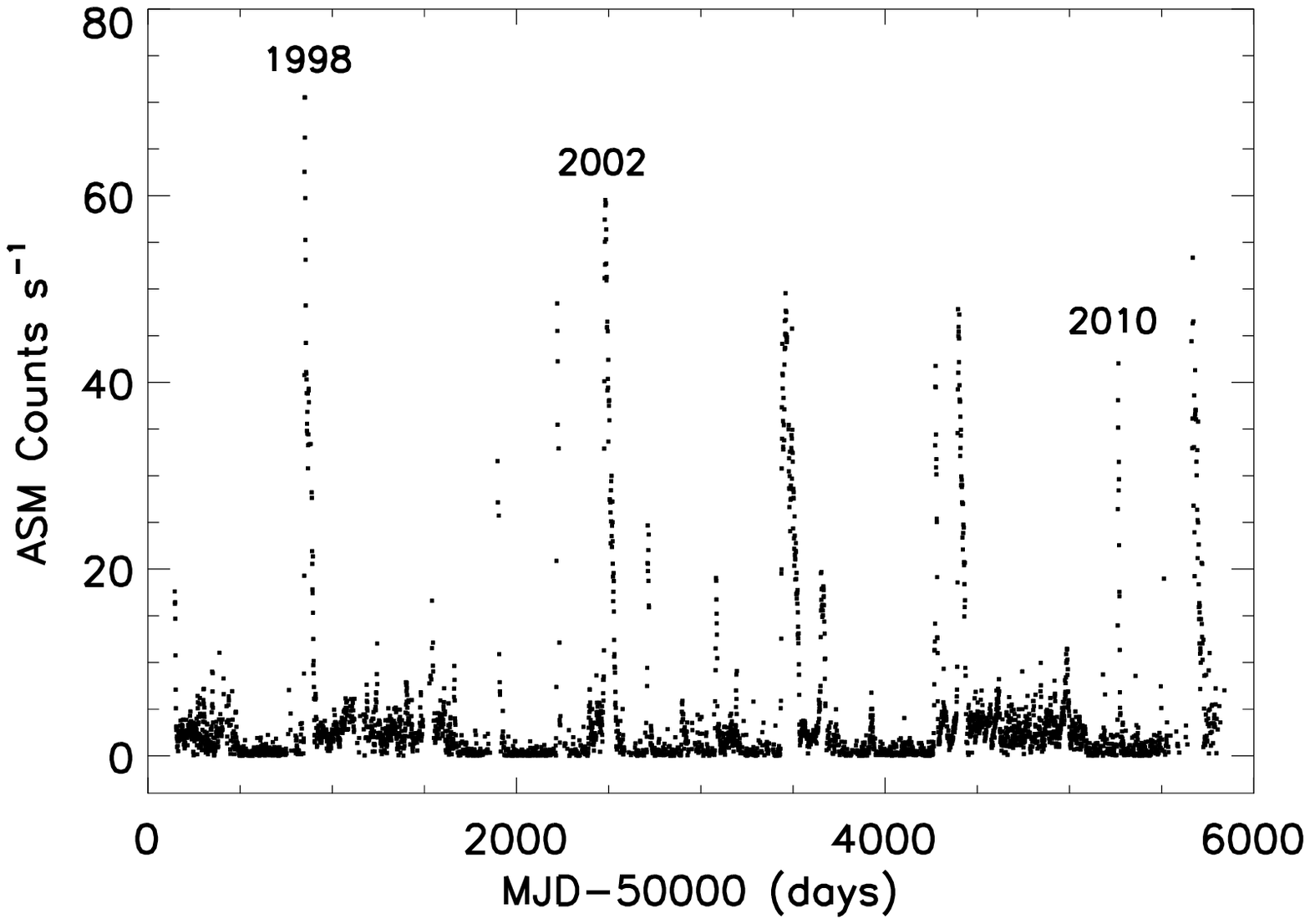}
\caption{Long-term  light curve of ASM for 4U 1608-52.}\label{fig1}
\end{center}
\end{figure*}

\begin{figure*}
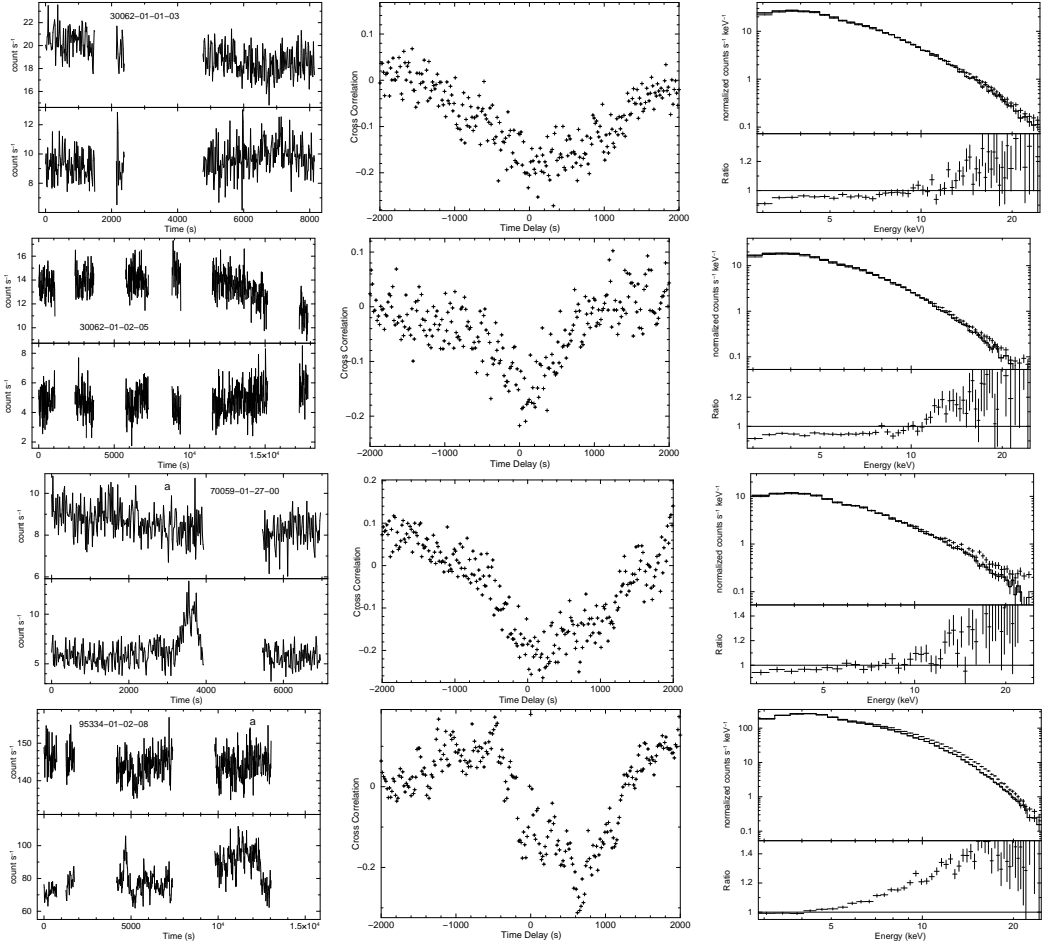

\begin{center}
\includegraphics[width=3.1cm,angle=270,clip]{fig2a.ps}
\includegraphics[width=3.1cm,angle=270,clip]{fig2b.ps}
\includegraphics[width=3.1cm,angle=270,clip]{fig2c.ps}
\includegraphics[width=3.1cm,angle=270,clip]{fig2d.ps}
\includegraphics[width=3.1cm,angle=270,clip]{fig2e.ps}
\includegraphics[width=3.1cm,angle=270,clip]{fig2f.ps}
\includegraphics[width=3.1cm,angle=270,clip]{fig2g.ps}
\includegraphics[width=3.1cm,angle=270,clip]{fig2h.ps}
\includegraphics[width=3.1cm,angle=270,clip]{fig2i.ps}
\includegraphics[width=3.1cm,angle=270,clip]{fig2j.ps}
\includegraphics[width=3.1cm,angle=270,clip]{fig2k.ps}
\includegraphics[width=3.1cm,angle=270,clip]{fig2l.ps}
\caption{From left to right: the lightcurves for which anti-correlation is detected, the corresponding cross-correlation functions, the X-ray spectra  for
hard and soft regions of the lightcurves. } \label{fig2}
\end{center}
\end{figure*}

\begin{figure*}
\begin{center}
\includegraphics[width=5.1cm,angle=0,clip]{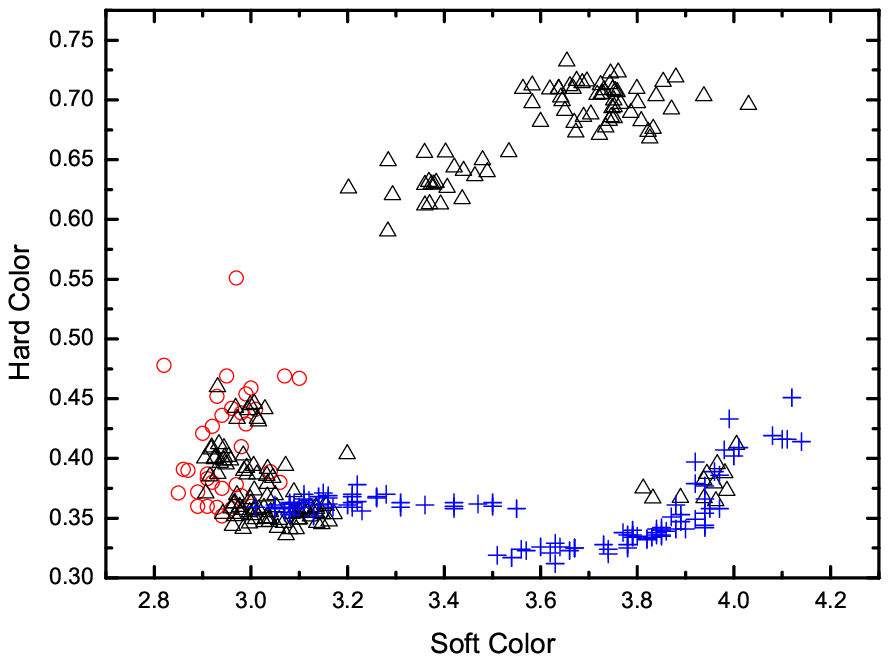}
\includegraphics[width=5.1cm,angle=0,clip]{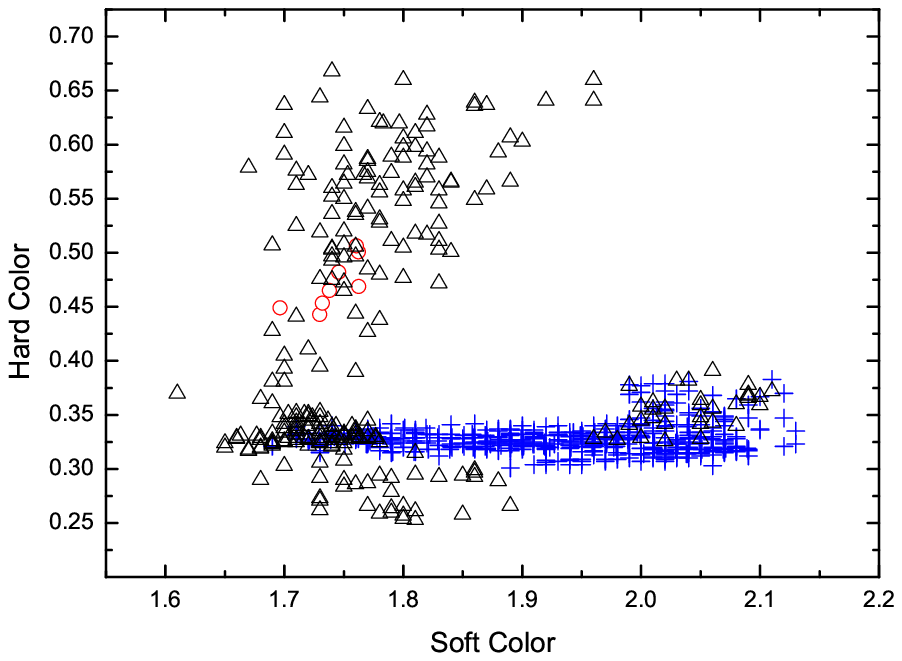}
\includegraphics[width=5.1cm,angle=0,clip]{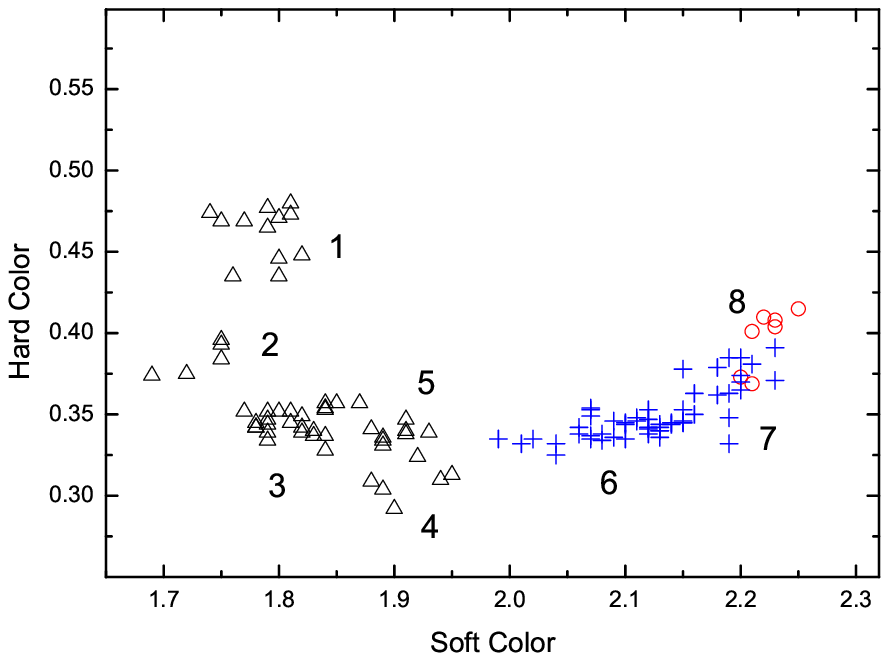}
\caption{From left to right: the CCDs of 1998, 2002 and 2010 outbursts.
The positive, ambiguous, and anti-correlations are represented
by blue crosses, black triangles, and red open circles, respectively.
}
\label{fig3}
\end{center}
\end{figure*}

\begin{figure*}
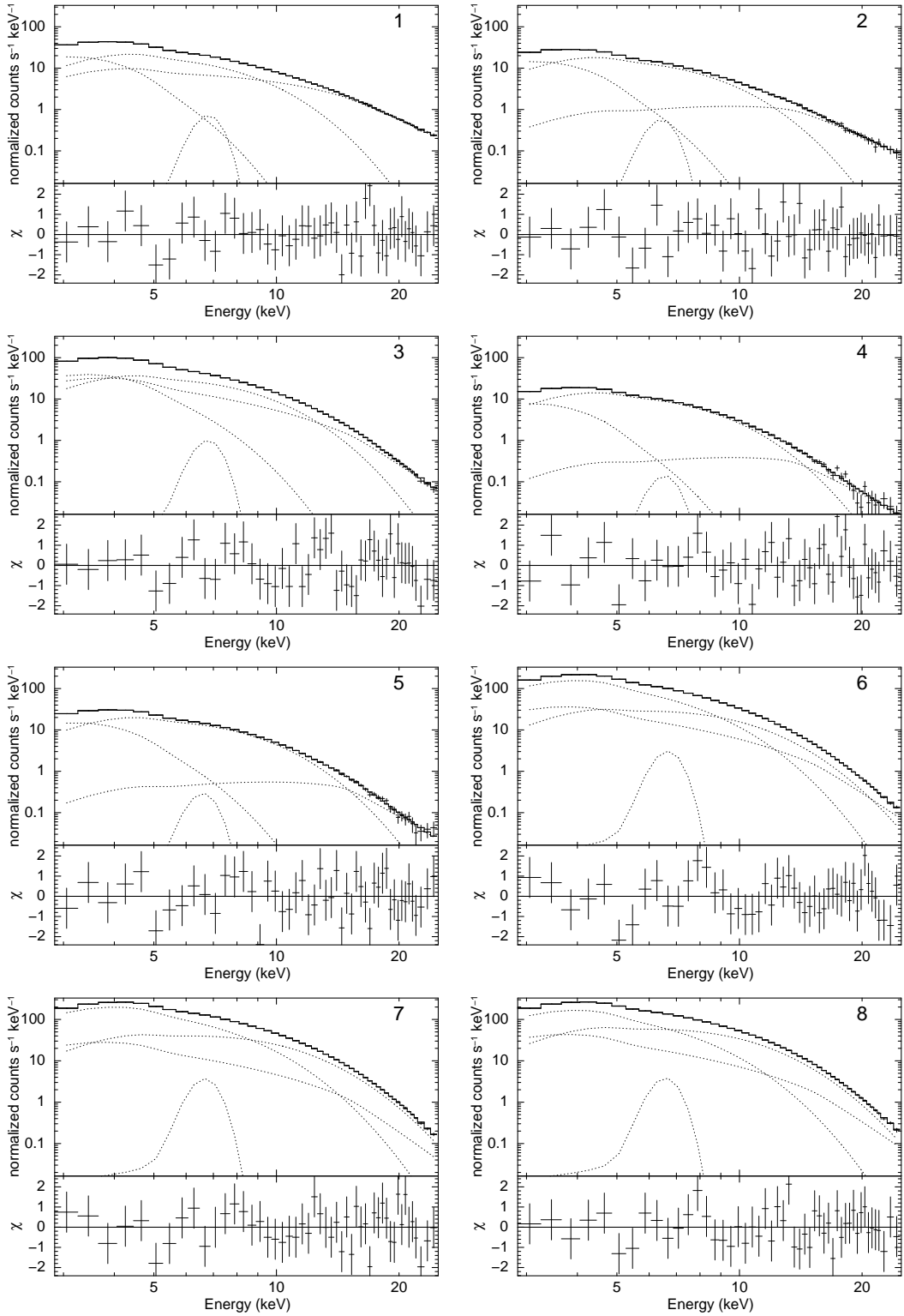

\begin{center}
\includegraphics[width=5.1cm,angle=270,clip]{fig4a.ps}
\includegraphics[width=5.1cm,angle=270,clip]{fig4b.ps}
\includegraphics[width=5.1cm,angle=270,clip]{fig4c.ps}
\includegraphics[width=5.1cm,angle=270,clip]{fig4d.ps}
\includegraphics[width=5.1cm,angle=270,clip]{fig4e.ps}
\includegraphics[width=5.1cm,angle=270,clip]{fig4f.ps}
\includegraphics[width=5.1cm,angle=270,clip]{fig4g.ps}
\includegraphics[width=5.1cm,angle=270,clip]{fig4h.ps}

\caption{The spectrum for each region is plotted along with its model components,  where the  $\Delta \chi$ of the
corresponding best fit is shown at each lower panel. }
\label{fig4}
\end{center}
\end{figure*}

\begin{deluxetable}{lccccc}
\tablewidth{0pt}
\tablecaption{Log of all observations ($>$ 2000 s) which show a positive correlation or ambiguous correlation}
\tablehead{
\colhead{ObsID}
& \colhead{Date}
& \colhead{Delay (s)}
& \colhead{CC}
& \colhead{Hardness Ratio}
}

\startdata

 \hline
&&1998 outburst&& \\

 \noalign{\smallskip}\hline\noalign{\smallskip}

& &  positive-correlation  & & \\
 \noalign{\smallskip}\hline\noalign{\smallskip}

30188-01-01-00     &  1998-02-03  & 30$\pm$87      &  0.41$\pm$0.02   & 0.361/3.467  \\
30188-01-04-00     &  1998-02-05  & 14$\pm$26      &  0.41$\pm$0.04   & 0.338/3.811  \\
30062-03-01-00     &  1998-02-06  & -3.0$\pm$7.7   &  0.40$\pm$0.06   & 0.338/3.840  \\
30062-03-01-02\_1   &  1998-02-06  & 17.3$\pm$19.0  &  0.26$\pm$0.04   & 0.388/3.938  \\
30062-03-01-02\_2   &  1998-02-06  & 11.1$\pm$12.1  &  0.34$\pm$0.03   & 0.409/4.058  \\
30062-03-01-02\_3   &  1998-02-06  & -8.4$\pm$20.5  &  0.27$\pm$0.04   & 0.365/3.936  \\
30062-03-01-03\_1   &  1998-02-06  & 18.1$\pm$19.0  &  0.44$\pm$0.04   & 0.330/3.798  \\
30062-03-01-03\_2   &  1998-02-06  & -27.0$\pm$28.1 &  0.31$\pm$0.08   & 0.326/3.625  \\
30502-01-10-00     &  1998-02-07  & -11$\pm$28     &  0.70$\pm$0.05   & 0.333/3.799  \\
30062-02-01-02\_2   &  1998-03-26  & 248$\pm$77     &  0.11$\pm$0.02   & 0.361/3.087  \\
30062-01-02-02     &  1998-04-04  & -101$\pm$236   &  0.14$\pm$0.03   & 0.364/3.135  \\
30062-01-02-00\_1   &  1998-04-05  &  46$\pm$53     &  0.25$\pm$0.03   & 0.362/3.241  \\
30062-01-02-00\_2   &  1998-04-05  & -7.5$\pm$17.9  &  0.20$\pm$0.05   & 0.368/3.207  \\
30062-01-02-00\_3   &  1998-04-05  & -5.1$\pm$12.0  &  0.27$\pm$0.07   & 0.365/3.142  \\
 \noalign{\smallskip}\hline\noalign{\smallskip}

  & &  ambiguous & & & \\

  \noalign{\smallskip}\hline\noalign{\smallskip}
30188-01-11-00     &  1998-02-08  &   \nodata        &  \nodata      &  0.382/3.958  \\
30062-02-01-01\_1   &  1998-03-25  &   \nodata        &  \nodata      &  0.389/3.037  \\
30062-02-01-01\_2   &  1998-03-25  &   \nodata        &  \nodata      &  0.400/2.948  \\
30062-02-01-01\_3   &  1998-03-25  &   \nodata        &  \nodata      &  0.403/2.929  \\
30062-02-01-01\_4   &  1998-03-25  &   \nodata        &  \nodata      &  0.389/2.945  \\
30062-02-01-02\_1   &  1998-03-26  &   \nodata        &  \nodata      &  0.359/3.130   \\
30062-01-01-00\_2   &  1998-03-27  &   \nodata        &  \nodata      &  0.355/3.097  \\
30062-01-01-01\_2   &  1998-03-28  &   \nodata        &  \nodata      &  0.354/3.002  \\
30062-01-01-02     &  1998-03-29  &   \nodata        &  \nodata      &  0.362/2.992  \\
30062-01-01-04\_1   &  1998-03-31  &   \nodata        &  \nodata      &  0.620/3.377  \\
30062-01-01-04\_2   &  1998-03-31  &   \nodata        &  \nodata      &  0.627/3.396  \\
30062-01-01-04\_3   &  1998-03-31  &   \nodata        &  \nodata      &  0.647/3.396  \\
30062-01-02-01     &  1998-04-03  &   \nodata        &  \nodata      &  0.441/2.993  \\
30062-01-02-03     &  1998-04-07  &   \nodata        &  \nodata      &  0.353/3.023   \\
30062-01-02-04\_1   &  1998-04-09  &   \nodata        &  \nodata      &  0.349/3.110  \\
30062-01-02-04\_2   &  1998-04-09  &   \nodata        &  \nodata      &  0.346/3.073   \\
30062-02-02-01     &  1998-04-20  &   \nodata        &  \nodata      &  0.690/3.713   \\
30062-02-02-00     &  1998-04-22  &   \nodata        &  \nodata      &  0.678/3.784  \\
30062-01-03-00\_1   &  1998-06-03  &   \nodata        &  \nodata      &  0.708/3.708  \\
30062-01-03-00\_2   &  1998-06-03  &   \nodata        &  \nodata      &  0.712/3.708  \\
30062-01-03-00\_3   &  1998-06-03  &   \nodata        &  \nodata      &  0.706/3.735  \\
30062-01-03-00\_4   &  1998-06-03  &   \nodata        &  \nodata      &  0.712/3.692  \\

 \noalign{\smallskip}\hline\noalign{\smallskip}

&&2002 outburst && \\
\noalign{\smallskip}\hline\noalign{\smallskip}

& &  positive-correlation  & & \\
 \noalign{\smallskip}\hline\noalign{\smallskip}

70058-01-15-00   & 2002-07-27 &  3.3$\pm$6.1    &  0.52$\pm$0.08  & 0.3200/2.106 \\
70059-01-03-01   & 2002-08-02 &  19$\pm$24      &  0.32$\pm$0.04  & 0.3139/2.026  \\
70059-01-03-02   & 2002-08-02 &  -12$\pm$12     &  0.43$\pm$0.03  & 0.3140/2.051  \\
70058-01-20-00   & 2002-08-03 &  -2.1$\pm$30.7  &  0.38$\pm$0.05  & 0.3159/2.056 \\
70059-01-04-00\_1 & 2002-08-04 &  -11$\pm$45     &  0.27$\pm$0.05  & 0.3115/1.960  \\
70059-01-04-00\_2 & 2002-08-04 &  15$\pm$18      &  0.52$\pm$0.02  & 0.3059/1.931  \\
70058-01-21-00   & 2002-08-05 &  9.4$\pm$10.1   &  0.60$\pm$0.02  & 0.3133/2.031  \\
70059-01-05-00\_1 & 2002-08-06 &  2.3$\pm$8.2    &  0.45$\pm$0.06  & 0.3153/2.035  \\
70059-01-05-00\_2 & 2002-08-06 &  -3.6$\pm$16.6  &  0.32$\pm$0.07  & 0.3178/2.061  \\
70058-01-23-01   & 2002-08-07 &  13$\pm$19      &  0.23$\pm$0.08  & 0.3206/1.954  \\
70058-01-23-00   & 2002-08-10 &  5.1$\pm$6.2    &  0.53$\pm$0.06  & 0.3190/2.007  \\
70059-01-07-00\_1 & 2002-08-11 &  -6.3$\pm$9.3   &  0.52$\pm$0.06  & 0.3211/2.032  \\
70059-01-07-00\_2 & 2002-08-11 &  0.7$\pm$9.3    &  0.56$\pm$0.03  & 0.3354/2.083  \\
70058-01-25-00   & 2002-08-12 &  -25$\pm$30     &  0.26$\pm$0.04  & 0.3192/1.897  \\
70059-01-08-00\_2 & 2002-08-13 &  4.3$\pm$9.6    &  0.44$\pm$0.06  & 0.3224/2.028  \\
70059-01-08-00\_4 & 2002-08-13 & -5.3$\pm$8.1    &  0.51$\pm$0.05  & 0.3278/2.041  \\
70058-01-24-00   & 2002-08-14 &  -13$\pm$15     &  0.26$\pm$0.04  & 0.3542/2.076  \\
70059-01-09-00\_1 & 2002-08-16 & 11$\pm$19       &  0.48$\pm$0.08  & 0.3293/2.048  \\
70059-01-09-00\_3 & 2002-08-16 & 1.4$\pm$8.6     &  0.25$\pm$0.08  & 0.3456/2.069  \\
70058-01-26-00   & 2002-08-17 &  8.0$\pm$16.5   &  0.32$\pm$0.04  & 0.3193/1.8005 \\
70059-01-10-00\_1 & 2002-08-18 &  26$\pm$42     &  0.19$\pm$0.04  & 0.3242/1.899  \\
70059-01-10-00\_2 & 2002-08-18 &  -20$\pm$32    &  0.24$\pm$0.05  & 0.3212/1.952  \\
70059-01-10-00\_3 & 2002-08-18 &  19$\pm$25     &  0.35$\pm$0.04  & 0.3390/2.051  \\
70058-01-27-00   & 2002-08-19 &  -16$\pm$17     &  0.34$\pm$0.04  & 0.3259/1.847  \\
70059-01-11-00\_1 & 2002-08-20 &  85$\pm$106    &  0.41$\pm$0.03  & 0.3230/1.838  \\
70059-01-11-00\_2 & 2002-08-20 &  -15$\pm$15    &  0.32$\pm$0.05  & 0.3223/1.916  \\
70059-01-11-00\_3 & 2002-08-20 &  2.2$\pm$14.4  &  0.59$\pm$0.03  & 0.3230/1.922  \\
70058-01-28-00   & 2002-08-21 &  -2.6$\pm$37.1  &  0.22$\pm$0.05  & 0.3243/1.909  \\
70059-01-12-00   & 2002-08-22 &  -7.7$\pm$14.0 &  0.31$\pm$0.05  & 0.3385/2.019  \\
70059-01-12-01\_1 & 2002-08-22 &  -6.7$\pm$18.7 &  0.25$\pm$0.07  & 0.3405/1.984  \\
70059-01-12-01\_2 & 2002-08-22 &  2.4$\pm$4.1   &  0.41$\pm$0.15  & 0.3757/2.003  \\
70058-01-29-00   & 2002-08-23 &  -4.0$\pm$24.5  &  0.18$\pm$0.05  & 0.3529/2.021  \\

70058-01-30-00   & 2002-08-25 &  0.5$\pm$4.7    &  0.43$\pm$0.15  & 0.3262/1.873  \\
70059-01-14-01   & 2002-08-26 &  -6.3$\pm$12.2 &  0.36$\pm$0.06  & 0.3303/1.949  \\
70059-01-14-00\_1 & 2002-08-26 &  -13$\pm$10    &  0.59$\pm$0.02  &  0.3348/2.017 \\

70058-01-31-00   & 2002-08-27 &  23$\pm$23     &  0.25$\pm$0.06  & 0.3309/1.884  \\
70059-01-15-00\_1 & 2002-08-28 &  -19$\pm$29    &  0.41$\pm$0.03  & 0.3413/1.996  \\
70059-01-15-00\_2 & 2002-08-28 &  -0.3$\pm$9.4  &  0.46$\pm$0.09  & 0.3276/1.977  \\
70059-01-15-00\_3 & 2002-08-28 &  9.5$\pm$9.8   &  0.22$\pm$0.05  & 0.3567/2.021  \\
70058-01-32-00   & 2002-08-29 &  20$\pm$13      &  0.38$\pm$0.06  & 0.3274/1.904  \\
70059-01-16-00\_1 & 2002-08-30 &  -1.0$\pm$32.0 &  0.25$\pm$0.03  & 0.3223/1.803  \\
70059-01-16-00\_2 & 2002-08-30 &  -7.2$\pm$13.1 &  0.26$\pm$0.13  & 0.3238/1.803  \\
70059-01-16-00\_3 & 2002-08-30 &  -16$\pm$24    &  0.19$\pm$0.06  & 0.3265/1.863  \\
70059-01-17-01   & 2002-09-01 &  47$\pm$51     &  0.23$\pm$0.03  & 0.3337/1.756  \\
70059-01-17-00   & 2002-09-01 &  23$\pm$23     &  0.20$\pm$0.05  & 0.3295/1.782  \\

70059-01-18-00\_2 & 2002-09-03 & -22$\pm$19     &  0.20$\pm$0.07  & 0.3299/1.841  \\
70058-01-35-00   & 2002-09-04 &  3.6$\pm$22.5   &  0.42$\pm$0.15  & 0.3290/1.910  \\
70059-01-19-00\_1 & 2002-09-05 &  -15$\pm$21    &  0.32$\pm$0.03  & 0.3282/1.756  \\
70059-01-19-00\_2 & 2002-09-05 &  4.3$\pm$15.2  &  0.31$\pm$0.10  & 0.3282/1.777  \\
70058-01-36-00   & 2002-09-06 &  -50$\pm$44     &  0.26$\pm$0.03  & 0.3341/1.756  \\
70059-01-21-00\_2 & 2002-09-09 & 1$\pm$115      &  0.15$\pm$0.03  & 0.3349/1.903  \\

70058-01-38-00   & 2002-09-10 &  15$\pm$29      &  0.23$\pm$0.06  & 0.3342/1.954  \\
70059-01-22-00\_1 & 2002-09-11 &  0.9$\pm$9.8   &  0.35$\pm$0.06  & 0.3338/1.888  \\
70059-01-22-00\_2 & 2002-09-11 &  22$\pm$56     &  0.36$\pm$0.03  & 0.3346/1.784  \\
70059-01-22-00\_3 & 2002-09-11 & 27$\pm$43      &  0.16$\pm$0.04  & 0.3317/1.737  \\

70059-03-01-00   & 2002-09-12 &  19$\pm$19     &  0.25$\pm$0.05  & 0.3270/1.708  \\
70059-01-23-00   & 2002-09-21 &  6.3$\pm$18.4  &  0.17$\pm$0.05  & 0.3217/1.773  \\

 \noalign{\smallskip}\hline\noalign{\smallskip}

  & &  ambiguous & & & \\

  \noalign{\smallskip}\hline\noalign{\smallskip}

70058-01-09-00   & 2002-05-30 &   \nodata        &  \nodata      &  0.6446/1.926  \\
70059-01-08-00\_1 & 2002-08-13 &   \nodata        &  \nodata      &  0.3528/2.078  \\
70059-01-09-00\_2 & 2002-08-16 &   \nodata        &  \nodata      &  0.3647/2.083  \\
70059-01-13-00   & 2002-08-24 &   \nodata        &  \nodata      &  0.3558/2.003  \\
70059-01-14-00\_2 & 2002-08-26 &   \nodata        &  \nodata      &  0.3412/2.024  \\
70059-01-20-00   & 2002-09-07 &   \nodata        &  \nodata      &  0.3468/1.714  \\
70058-01-37-00   & 2002-09-08 &   \nodata        &  \nodata      &  0.3404/1.733  \\
70059-03-01-01   & 2002-09-11 &   \nodata        &  \nodata      &  0.3336/1.705  \\
70059-03-02-00   & 2002-09-14 &   \nodata        &  \nodata      &  0.3296/1.744  \\
70059-03-02-02   & 2002-09-14 &   \nodata        &  \nodata      &  0.3286/1.741  \\
70059-03-02-07   & 2002-09-15 &   \nodata        &  \nodata      &  0.3271/1.707  \\
70069-01-01-01   & 2002-09-20 &   \nodata        &  \nodata      &  0.3250/1.751  \\
70059-01-23-00   & 2002-09-21 &   \nodata        &  \nodata      &  0.3229/1.726  \\
70069-01-01-03   & 2002-09-24 &   \nodata        &  \nodata      &  0.3711/1.682  \\
70069-01-01-00   & 2002-09-25 &   \nodata        &  \nodata      &  0.4922/1.746  \\
70069-01-01-04   & 2002-09-26 &   \nodata        &  \nodata      &  0.5884/1.798  \\
70069-01-02-00   & 2002-09-27 &   \nodata        &  \nodata      &  0.6174/1.729  \\
70059-01-26-00   & 2002-09-29 &   \nodata        &  \nodata      &  0.3156/1.698  \\
70069-01-03-03   & 2002-10-02 &   \nodata        &  \nodata      &  0.4932/1.793  \\
70059-01-28-00   & 2002-10-03 &   \nodata        &  \nodata      &  0.5097/1.770  \\
70069-01-03-00   & 2002-10-05 &   \nodata        &  \nodata      &  0.5676/1.811  \\
70059-01-24-00   & 2002-10-05 &   \nodata        &  \nodata      &  0.5640/1.778  \\
70069-01-03-04   & 2002-10-07 &   \nodata        &  \nodata      &  0.4357/1.740  \\
70069-01-03-07   & 2002-10-08 &   \nodata        &  \nodata      &  0.4000/1.716  \\
70059-01-29-00   & 2002-10-09 &   \nodata        &  \nodata      &  0.2754/1.770  \\
70069-01-03-08   & 2002-10-10 &   \nodata        &  \nodata      &  0.2943/1.850  \\
70069-01-04-00   & 2002-10-11 &   \nodata        &  \nodata      &  0.2591/1.786  \\
70069-01-04-01   & 2002-10-12 &   \nodata        &  \nodata      &  0.5530/1.807  \\
70069-01-04-07   & 2002-10-12 &   \nodata        &  \nodata      &  0.5795/1.827  \\
70069-01-04-02   & 2002-10-13 &   \nodata        &  \nodata      &  0.6205/1.849  \\
70069-01-04-03   & 2002-10-14 &   \nodata        &  \nodata      &  0.6135/1.768  \\
70069-01-07-05   & 2002-11-17 &   \nodata        &  \nodata      &  0.3381/1.859  \\

\noalign{\smallskip}\hline\noalign{\smallskip}
&&2010 outburst&& \\
\noalign{\smallskip}\hline\noalign{\smallskip}

& &  positive-correlation  & & \\
 \noalign{\smallskip}\hline\noalign{\smallskip}

  95334-01-02-04   & 2010-03-07  &  -17$\pm$63      & 0.29$\pm$0.06 & 0.3404/2.102  \\
  95334-01-02-14   & 2010-03-07  &  11$\pm$12       & 0.38$\pm$0.05 & 0.3412/2.134 \\
  95334-01-02-08\_1 & 2010-03-08  &  -9.5$\pm$22.0   & 0.29$\pm$0.06 & 0.3757/2.187  \\
  95334-01-02-01   & 2010-03-09  &  -0.3$\pm$11.9   & 0.29$\pm$0.05 & 0.3330/2.037  \\
  95334-01-02-11   & 2010-03-10  &  18$\pm$33       & 0.38$\pm$0.04 & 0.3694/2.203  \\
  95334-01-03-00   & 2010-03-12  &  8.7$\pm$12.1    & 0.40$\pm$0.05 & 0.3444/2.123 \\
  95334-01-03-01   & 2010-03-13  &  -76$\pm$93      & 0.14$\pm$0.04 & 0.3496/2.082 \\
\noalign{\smallskip}\hline\noalign{\smallskip}

  & &  ambiguous & & & \\

  \noalign{\smallskip}\hline\noalign{\smallskip}
 95334-01-01-00   & 2010-03-04  &  \nodata          &  \nodata      & 0.3098/1.910 \\
 95334-01-02-03   & 2010-03-05  &  \nodata          &  \nodata      & 0.3378/1.884 \\
 95334-01-03-03   & 2010-03-15  &  \nodata          &  \nodata      & 0.3535/1.838 \\
 95334-01-03-04   & 2010-03-16  &  \nodata          &  \nodata      & 0.3492/1.790 \\
 95334-01-03-05   & 2010-03-17  &  \nodata          &  \nodata      & 0.3411/1.801 \\
 95334-01-03-06   & 2010-03-18  &  \nodata          &  \nodata      & 0.4626/1.789 \\
 95334-01-04-01   & 2010-03-20  &  \nodata          &  \nodata      & 0.3860/1.742 \\

\enddata \\
\footnotesize {We label the segments ($>$ 2000 s) of the light curves by  $1$,
$2$, and so on. }
\end{deluxetable}

\begin{table*}
\footnotesize
\caption{\bf ~~Log of observations where the anti-correlations  are detected}
\scriptsize{} \label{table:2}
\newcommand{\m}{\hphantom{$-$}}
\newcommand{\cc}[1]{\multicolumn{1}{c}{#1}}
\renewcommand{\tabcolsep}{0.6pc} 
\renewcommand{\arraystretch}{1.2} 
\medskip
\begin{center}
\begin{tabular}{c c c c c c c c c}
\hline
      ObsID& Date& Location& Delay (s)& CC& Hardness Ratio & Pivoting \\
\hline

  30062-01-01-03 & 1998-03-30   & LLB &     136$\pm$136     & -0.18$\pm$0.01        &  0.4468/2.978  & $\sim$8.5 \\
  30062-01-02-05 & 1998-04-11   & LLB &      -100$\pm$102   &  -0.15$\pm$0.02      &  0.3866/2.943  & $\sim$10 \\
  70059-01-27-00 & 2002-10-01   & LLB &       168$\pm$32    &  -0.22$\pm$0.01       &  0.4712/1.741  & $\sim$6\\
  95334-01-02-08\_2 & 2010-03-08  & UB &     628$\pm$43     & -0.18$\pm$0.02        &  0.4000/2.223  & $\sim$4\\

\hline
\end{tabular}
\end{center}
\footnotesize
\end{table*}

\begin{table*}
\footnotesize
\caption{\bf ~~The fitting parameters of the spectra of 2010 outburst}
\scriptsize{} \label{table:3}
\newcommand{\m}{\hphantom{$-$}}
\newcommand{\cc}[1]{\multicolumn{1}{c}{#1}}
\renewcommand{\tabcolsep}{0.2pc} 
\renewcommand{\arraystretch}{1.3} 
\medskip
\begin{center}
\begin{tabular}{c c c c c c c c c c }
\hline
Parameters & $kT_{in}$(keV)    &  $N_{disk}$       & $kT_{bb}$(keV)    & $N_{bb}$    & $Flux_{diskbb}/Flux_{total}$ &$Flux_{bb}/Flux_{total}$&$Flux_{BPL}/Flux_{total}$ &$\chi^2$(d.o.f) \\
\cline{5-6}

\hline
 1   &  $0.68_{-0.06}^{+0.07}$ &  $671_{-234}^{+338}$  & $1.49_{-0.05}^{+0.04}$  & $0.014_{-0.002}^{+0.002}$ &  0.44/2.56 (17.2\%) & 1.00/2.56 (39.1\%) &1.11/2.56(43.4\%) &  1.01(47)  \\

 2   &  $0.69_{-0.06}^{+0.06}$ &  $460_{-150}^{+240}$  & $1.53_{-0.06}^{+0.05}$  & $0.012_{-0.001}^{+0.001}$ &  0.34/1.50 (22.7\%) & 0.87/1.50 (58.0\%)&0.29/1.50(19.3\%) &  0.75(46)  \\

 3   &  $0.90_{-0.06}^{+0.06}$ &  $272_{-53}^{+147}$  & $1.68_{-0.06}^{+0.06}$  & $0.027_{-0.002}^{+0.002}$ &  1.04/4.73 (22.0\%) & 1.97/4.73 (41.6\%)&1.70/4.73(35.9\%) & 1.12(46)  \\

 4   &  $0.66_{-0.06}^{+0.06}$ &  $333_{-127}^{+221}$  & $1.54_{-0.04}^{+0.04}$ & $0.010_{-0.001}^{+0.001}$ &  0.18/0.94 (19.0\%) & 0.72/0.94 (72.3\%)&0.08/0.94(8.5\%) &  1.08(41)  \\

 5   &  $0.76_{-0.05}^{+0.05}$ &  $279_{-69}^{+115}$  & $1.64_{-0.05}^{+0.03}$  & $0.014_{-0.001}^{+0.001}$ &  0.37/1.53 (24.2\%) & 1.02/1.53 (66.7\%)&0.14/1.53(9.2\%) &  0.99(40)  \\

 6   &  $1.64_{-0.08}^{+0.13}$ &  $68.2_{-16.2}^{+14.6}$  & $2.25_{-0.12}^{+0.11}$  & $0.034_{-0.009}^{+0.007}$ &  5.87/10.5 (55.9\%) & 2.67/10.5 (25.4\%)&1.92/10.5(18.3\%) &  0.89(40)  \\
 7   &  $1.71_{-0.10}^{+0.11}$ &  $75.1_{-17.9}^{+20.0}$  & $2.42_{-0.10}^{+0.13}$  & $0.053_{-0.012}^{+0.009}$ &  7.66/13.28 (57.7\%) & 4.11/13.28 (30.9\%)&1.47/13.28(11.1\%) &  0.92(40)  \\
 8   &  $1.62_{-0.11}^{+0.23}$ &  $77.2_{-31.0}^{+22.4}$  & $2.33_{-0.08}^{+0.16}$  & $0.073_{-0.019}^{+0.010}$ &  6.18/14.23 (43.4\%) &5.72/14.23 (40.2\%)&2.29/14.23(16.1\%) &  0.76(40)  \\

\hline

\end{tabular}
\end{center}
\footnotesize {The flux is calculated in the energy band 2.5-25 keV, and is  in unit of 10$^{-9}$ ergs cm$^{-2}$ s$^{-1}$. Errors are quoted at a
90\% confidence level.  The letters 1, 2, 3 and so on are labeled as the
segments in the light curve.}
\end{table*}


\begin{thebibliography}{}




\bibitem[Agrawal \& Sreekumar(2003)]{agr03} Agrawal, V.~K., \& Sreekumar, P. 2003, \mnras, 346, 933
\bibitem[Arnaud(1996)]{arn96} Arnaud, K.~A. 1996, Astronomical Data Analysis Software and Systems V, 101, 17

\bibitem[Belloni(2010)]{bel10} Belloni, T.~M. 2010, Lecture Notes in Physics, Berlin Springer Verlag, 794, 53
\bibitem[Berger et al.(1996)]{ber96} Berger, M., van der
Klis, M., van Paradijs, J., et al.\ 1996, \apjl, 469, L13



\bibitem[B{\"o}ttcher \& Liang(1999)]{bot99} B{\"o}ttcher, M., \& Liang, E.~P. 1999, \apjl, 511, L37


\bibitem[Brinkman et al.(1974)]{bri74} Brinkman, A.~C., Parsignault, D.~R., Schreier, E., et al.\ 1974, \apj, 188, 603

\bibitem[Casares et al.(2006)]{cas06} Casares, J.,
Cornelisse, R., Steeghs, D., et al.\ 2006, \mnras, 373, 1235

\bibitem[Chen et al.(2006)]{che06} Chen, X., Zhang, S.~N.,
\& Ding, G.~Q.\ 2006, \apj, 650, 299
\bibitem[Choudhury \& Rao(2004)]{cho04} Choudhury, M., \& Rao, A.~R. 2004, \apjl, 616, L143
\bibitem[Choudhury et al.(2005)]{cho05} Choudhury, M., Rao,
A.~R., Dasgupta, S., et al.\ 2005, \apj, 631, 1072


\bibitem[Church \& Balucinska-Church(1995)]{chu95} Church, M.~J., \& Balucinska-Church, M. 1995, \aap, 300, 441


\bibitem[de Avellar et al.(2013)]{ave13} de Avellar,
M.~G.~B., M{\'e}ndez, M., Sanna, A.,
\& Horvath, J.~E.\ 2013, \mnras, 433, 3453

\bibitem[Di Salvo et al.(2000)]{dis00} Di Salvo, T., Iaria,
R., Burderi, L., \& Robba, N.~R.\ 2000, \apj, 542, 1034



\bibitem[Di Salvo et al.(2001)]{dis01} Di Salvo, T.,
M{\'e}ndez, M., van der Klis, M., Ford, E.,
\& Robba, N.~R.\ 2001, \apj, 546, 1107


\bibitem[di Salvo et al.(2009)]{dis09} di Salvo, T., D'A{\'{\i}}, A., Iaria, R., et al 2009, \mnras, 398, 2022
\bibitem[Done \& Gierli{\'n}ski(2006)]{don06} Done, C., \& Gierli{\'n}ski, M. 2006, \mnras, 367, 659
\bibitem[Done et al.(2007)]{don07} Done, C., Gierli{\'n}ski, M., \& Kubota, A.\ 2007, \aapr, 15, 1

\bibitem[Done \& Diaz Trigo(2010)]{don10} Done, C., \& Diaz Trigo, M. 2010, \mnras, 407, 2287
\bibitem[Esin et al.(1997)]{esi97} Esin, A.~A., McClintock,
J.~E., \& Narayan, R.\ 1997, \apj, 489, 865


\bibitem[Ford et al.(2000)]{for00} Ford, E.~C., van der Klis,
M., M{\'e}ndez, M., et al.\ 2000, \apj, 537, 368


\bibitem[Gierli{\'n}ski \& Done(2002)]{gie02} Gierli{\'n}ski, M., \& Done, C.\ 2002, \mnras, 337, 1373
\bibitem[G{\"u}ver et al.(2010)]{guv10} G{\"u}ver, T.,
{\"O}zel, F., Cabrera-Lavers, A., \& Wroblewski, P.\ 2010, \apj, 712, 964

\bibitem[Hasinger(1987)]{has87} Hasinger, G. 1987, IAU Symp.~125: The Origin and Evolution of Neutron Stars, 125, 333
\bibitem[Hasinger \& van der Klis(1989)]{has89} Hasinger, G., \& van der Klis, M. 1989, \aap, 225, 79
\bibitem[Hasinger et al.(1990)]{hasetal90} Hasinger, G., van der Klis, M., Ebisawa, K., Dotani, T., \& Mitsuda, K.\ 1990, \aap, 235, 131



\bibitem[Homan et al.(2010)]{hom10} Homan, J., van der Klis,
M., Fridriksson, J.~K., et al.\ 2010, \apj, 719, 201


\bibitem[Jackson et al.(2009)]{jac09} Jackson, N.~K., Church, M.~J., \& Ba{\l}uci{\'n}ska-Church, M. 2009, \aap, 494, 1059

\bibitem[Jahoda et al.(2006)]{jah06} Jahoda, K., Markwardt, C.~B., Radeva, Y., et al. 2006, \apjs, 163, 401
\bibitem[Jonker et al.(2000)]{jon00} Jonker, P.~G., van der
Klis, M., Wijnands, R., et al.\ 2000, \apj, 537, 374




\bibitem[Laurent \& Titarchuk(2007)]{lau07} Laurent, P., \& Titarchuk, L. 2007, \apj, 656, 1056

\bibitem[Lei et al.(2008)]{lei08} Lei, Y.~J., Qu, J.~L.,
Song, L.~M., et al.\ 2008, \apj, 677, 461


\bibitem[Lei et al.(2013b)]{lei13b} Lei, Y.~J., Zhang, H.~T.,
Luo, A.~L., \& Zhao, Y.~H.\ 2013, IAU Symposium, 290, 249


\bibitem[Lei et al.(2013a)]{lei13a} Lei, Y.-J., Zhang, H.-T.,
Zhang, C.-M., et al.\ 2013, \aj, 146, 60


\bibitem[Lin et al.(2007)]{lin07} Lin, D., Remillard, R.~A., \& Homan, J. 2007, \apj, 667, 1073

\bibitem[Lin et al.(2009)]{lin09} Lin, D., Remillard, R.~A.,
\& Homan, J.\ 2009, \apj, 696, 1257


\bibitem[Liu et al.(2007)]{liu07} Liu, B.~F., Taam, R.~E.,
Meyer-Hofmeister, E., \& Meyer, F.\ 2007, \apj, 671, 695

\bibitem[Markoff \& Nowak(2004)]{mar04} Markoff, S., \& Nowak, M.~A. 2004, \apj, 609, 972

\bibitem[McClintock \& Remillard(2006)]{mcc06} McClintock, J.~E., \& Remillard, R.~A. 2006, Compact stellar X-ray sources, 157
\bibitem[Mendez et al.(1997)]{men97} Mendez, M.,  et al. 1997, \apjl, 485, L37
\bibitem[Mendez et al.(1998)]{men98} Mendez, M., van der
Klis, M., Wijnands, R., et al.\ 1998, \apjl, 505, L23


\bibitem[Meyer et al.(2007)]{mey07} Meyer, F., Liu, B.F., \& Meyer-Hofmeister, E.\ 2007, \aap, 463, 1
\bibitem[Miller et al.(2006)]{mil06} Miller, J.~M.,  et al. 2006, \apj, 653, 525
\bibitem[Migliari et al.(2003)]{mig03} Migliari, S., Fender,
R.~P., Rupen, M., et al.\ 2003, \mnras, 342, L67

\bibitem[Mitsuda et al.(1989)]{mit89} Mitsuda, K., Inoue, H., Nakamura, N., \& Tanaka, Y. 1989, \pasj, 41, 97

\bibitem[Nowak et al.(1999)]{now99} Nowak, M.~A., Wilms, J.,
\& Dove, J.~B.\ 1999, \apj, 517, 355



\bibitem[O'Brien et al.(2004)]{obr04} O'Brien, K., Horne, K.,
Gomer, R.~H., Oke, J.~B., \& van der Klis, M.\ 2004, \mnras, 350, 587


\bibitem[Penninx et
al.(1989)]{pen89} Penninx, W., Damen, E., van Paradijs, J., Tan, J., \& Lewin, W.~H.~G.\ 1989, \aap, 208, 146
\bibitem[Popham \& Sunyaev(2001)]{pop01} Popham, R., \& Sunyaev, R 2001, AIP Conf.~Proc.~599: X-ray Astronomy: Stellar Endpoints,
AGN, and the Diffuse X-ray Background, 599, 870
\bibitem[Qu et al.(2001)]{qu01} Qu, J.~L., Yu, W., \& Li, T.~P. 2001, \apj, 555, 7
\bibitem[Remillard(2005)]{rem05} Remillard, R.~A. 2005,
in AIP Conf. Proc. 797, Interacting Binaries: Accretion,
Evolution, and Outcomes, ed. L. Burderi et al. (Melville, NY: AIP),231
\bibitem[Remillard \& McClintock(2006)]{rem06} Remillard, R.~A., \& McClintock, J.~E. 2006, \araa, 44, 49
\bibitem[Rykoff et al.(2007)]{ryk07} Rykoff, E.~S., Miller,
J.~M., Steeghs, D., \& Torres, M.~A.~P.\ 2007, \apj, 666, 1129


\bibitem[Schafer(1991)]{sch91} Schafer, R.~A. 1991, XSPEC,
an X-ray spectral fitting package: version 2 of the user's guide (ESA TM,
1013-7076, 09; Paris: European Space Agency)


\bibitem[Sriram et al.(2007)]{sri07} Sriram, K., Agrawal, V. K., Pendharkar, Jayant K., Rao, A. R. 2007, \apj, 661, 1055
\bibitem[Sriram et al.(2009)]{sri09} Sriram, Kandulapati; Kiron, Yellapragada Ravi; Vivekananda Rao, Pasagada 2009, RAA, 9, 901
\bibitem[Sriram et al.(2010)]{sri10} Sriram, K., Rao, A. R., Choi, C. S. 2010, \apj, 725, 1317
\bibitem[Sriram et al.(2012)]{sri12} Sriram, K., Choi, C. S., Rao, A. R. 2012, \apjs, 200, 16
\bibitem[Stella(1990)]{ste90} Stella, L. 1990, \nat, 344, 747

\bibitem[van der Klis(2000)]{van00} van der Klis, M. 2000, ARA\&A, 38, 717
\bibitem[van der Klis(2006)]{van06}
van der Klis, M. 2006, 
 in Compact stellar X-ray sources, W.H.G. Lewin \& M. van
der Klis  (Cambridge: Cambridge University Press) 39

\bibitem[van Paradijs
\& van der Klis(1994)]{van94} van Paradijs, J., \& van der Klis, M. 1994, \aap, 281, L17

\bibitem[van Straaten et al.(2003)]{van03} van Straaten, S.,
van der Klis, M., \& M{\'e}ndez, M.\ 2003, \apj, 596, 1155




\bibitem[Vaughan et al.(1997)]{vau97} Vaughan, B.~A., van der
Klis, M., Mendez, M., et al.\ 1997, \apjl, 483, L115
\bibitem[Vaughan et al.(1998)]{vau98} Vaughan, B.~A., van der Klis, M., M{\'e}ndez, M., et al.\ 1998, \apjl, 509, L145





\bibitem[Weisskopf et al.(1975)]{wei75} Weisskopf, M.~C.,
Kahn, S.~M., \& Sutherland, P.~G. 1975, \apjl, 199, L147

\bibitem[Weng \& Zhang(2011)]{wen11} Weng, S.S., \& Zhang, S.N.  2011, \apj, 739, 42
\bibitem[White et al.(1988)]{whi88} White, N.~E., Stella, L., \& Parmar, A.~N. 1988, \apj, 324, 363

\bibitem[Zhang(2007)]{zha07} Zhang, C.\ 2007, Advances in Space Research, 40, 1480


\end{thebibliography}
\end{document}